
%

\input phyzzx
\tolerance=5000
\sequentialequations
\overfullrule=0pt
\nopubblock
\PHYSREV
\doublespace
\def\section#1{\goodbreak\bigskip\goodbreak
	\centerline{{\bf #1}}\nobreak\medskip\nobreak\par}

\def\half{{1\over2}}
\def\a{\alpha}
\def\eps{\epsilon}
\def\ga{\gamma}
\def\la{\lambda}
\def\bfr{{\bf r}}
\def\p{{\bf P}}
\def\si{\sigma}
\def\beneathrel#1\under#2{\mathrel{\mathop{#2}\limits_{#1}}}
\def\gtwid{\mathrel{\raise.3ex\hbox{$>$\kern-.75em\lower1ex\hbox{$\sim$}}}}

\gdef\journal#1, #2, #3, 1#4#5#6{               
    {\sl #1~}{\bf #2}, #3 (1#4#5#6)}            
\def\mpl{\journal Mod. Phys. Lett., }
\def\np{\journal Nucl. Phys., }
\def\pl{\journal Phys. Lett., }
\def\prl{\journal Phys. Rev. Lett., }
\def\pr{\journal Phys. Rev., }
\def\prd{\journal Phys. Rev. D, }
\def\prb{\journal Phys. Rev. B, }
\def\ijmp{\journal Int. Jour. Mod. Phys., }

\line{\hfill UdeM-LPN-TH-94-188}
\line{\hfill hep-th/9405027}
\titlepage
\title{Parity violation, anyon scattering and the mean field approximation}
\author{Didier Caenepeel and Richard MacKenzie}
\vskip.2cm
\centerline{{\sl Laboratoire de Physique Nucl\'eaire}}
\centerline{{\sl Universit\'e de Montr\'eal}}
\centerline{{\sl Montr\'eal, Qu\'e, H3C 3J7}}
\centerline{{\sl Canada}}

\abstract{
Some general features of the scattering of boson-based anyons with an
added non-statistical interaction are discussed. Periodicity requirements
of the phase shifts are derived, and used to illustrate the danger inherent
in separating these phase shifts into the well-known pure Aharanov-Bohm
phase shifts, and an additional set which arise due to the interaction. It
is proven that the added phase shifts, although due to the non-statistical
interaction, necessarily change as the statistical parameter is varied,
keeping the interaction fixed. A hard-disk interaction provides a concrete
illustration of these general ideas.

In the latter part of the paper, scattering with an additional hard-disk
interaction is studied in detail, with an eye towards providing a criterion
for the validity of the mean-field approximation for anyons, which is the
first step in virtually any treatment of this system. We find, consistent
with previous work, that the approximation is justified if the statistical
interaction is weak, and that it must be more weak for boson-based than for
fermion-based anyons.
}
\endpage

\doublespace

\REF\leimyr{J.M. Leinaas and J. Myrheim,
\journal{Nuo. Cim.}, B37, 1, 1977.}

\REF\wilczekone{
F. Wilczek, \prl 48, 1144, 1982.}

\REF\wilczekthree{F. Wilczek, \prl 49, 957, 1982.}

\REF\reviews{For references and review, see, \eg, F. Wilczek, {\sl Fractional
statistics and anyon superconductivity} (World Scientific, 1990); A.P.
Balachandran, E. Ercolessi, G. Morandi, A.M. Srivastava, {\sl Hubbard Model
and Anyon Superconductivity} (World Scientific, 1990).}

\REF\aroschwilzee{
D.P. Arovas, \etal, \np B251 [FS13], 117, 1985.}

\REF\chewilwithal{Y-H. Chen, F. Wilczek,
E. Witten and B. Halperin, \ijmp B3, 1001, 1989.}

\REF\trug{C. Trugenberger, \prd 45, 3807, 1992.}

\REF\zee{A. Zee, ``Semionics: a Theory of High Temperature
Superconductivity,'' in {\sl High Temperature Superconductivity}, eds: K.
Bedell, D. Pines and J.R. Schrieffer (Addison-Wesley, 1990).}

\REF\mccmac{J. McCabe and R. MacKenzie, \mpl A8, 1909, 1993.}

\REF\caemac{D. Caenepeel and R. MacKenzie, \mpl A8, 1909, 1993.}

\REF\wenzee{X.G. Wen and A.Zee, \prb 41, 240, 1990.}

\REF\marwil{J. March-Russell and F. Wilczek, \prl 61, 2066, 1988.}

\REF\kiewei{ K. Kiers and N. Weiss, \prd 49, 2081, 1994.}

\REF\suzsribhalaw{A. Suzuki, M.K. Srivastava, R.K. Bhaduri
and J. Law, \prb 44, 10731, 1991.}

\REF\hagen{C.R. Hagen, \prd 41, 2015, 1990.}

\REF\ahaboh{Y. Aharonov and D. Bohm, \pr 115, 485, 1959.}

\REF\selfadj{
M.~Reed and B.~Simon,
{\sl Methods of Modern Mathematical Physics, vols.~1 and 2}
(Academic Press, N.Y., 1972);
P. de Sousa Gerbert, \prd 40, 1346, 1989;
P. de Sousa Gerbert and R. Jackiw, 
\journal Comm. Math. Phys., 124, 229, 1989.}

\REF\pedest{
K. Hwang, {\sl Quarks, Leptons and Gauge Fields}
(World Scientific, 1982);
R. Jackiw, "Delta-function potential in two- and
three-dimensional quantum mechanics", in M.A.B\'{e}g memorial
volume, A.Ali and P.Hoodbhoy, eds. (World Scientic, Singapore,
1991);
R. MacKenzie, P.K. Panigrahi, M.B. Paranjape, S. Sakhi, to be published,
{\sl Theoretical and Experimental Physics}.}

\REF\regular{
Hagen has recently emphasized the utility of using regularized versions of
the AB scattering problem; see C.R. Hagen, \ijmp A6, 3119, 1991. See also
J. Grundberg, \etal, \mpl B5, 539, 1991; S. Artz, \etal, \pl B267, 389,
1991.}

It is well known that in two space dimensions the possibility of fractional
statistics exists [\leimyr]. The simplest way of describing these
fractional-statistics particles,
anyons, is to start with conventional particles and attach fictitious
``statistical'' point charges and fluxes to the particles
[\wilczekone,\wilczekthree].\footnote{\hbox{$^\dagger$}}{In this article, we
consider only electrically neutral particles; thus any reference to charge,
flux, \etc, is understood to refer to the statistical quantities.}
The resulting Aharonov-Bohm (AB) interaction mimics
a change in statistics; for example, adopting a path-integral point of
view, a path involving a winding of one particle around another acquires an
additional contribution to its action from the statistical interaction
proportional to the change in relative angle, independent of the details of
the path. 

Depending on whether one starts from bosons or fermions, the
strength of the statistical interaction measures the departure of the
statistics from the ``base'' statistics. Evidently, one has a periodicity
requirement: starting from bosons, for example, one eventually
reaches a statistical interaction which transmutes them into fermions; one
requires that the dynamics of bosons transmuted to fermions agree with that
of the fermions themselves, all else being equal.

Anyons have been studied in great detail, mostly due to applications in the
fractional quantum Hall effect and to the prospect of a new mechanism of
superconductivity [\reviews]. Since the statistical interaction is rather
difficult to handle, the starting point is usually a ``mean-field
approximation'' (MFA), which consists in replacing the
flux carried by each anyon by a uniform magnetic field with the same
average value [\aroschwilzee]. 
This reduces the many-anyon problem to a system of many
conventional particles in a uniform magnetic field plus an interaction
which is a sum of the conventional (\eg, Coulomb) interaction between anyons
and an interaction term due to the MFA (equal to the
point magnetic fields of the individual anyons {\sl minus} the mean field).

At an intuitive level, since a sum of $\delta$-functions is exceedingly {\sl
non-uniform}, the validity of the MFA merits some
scrutiny. Several arguments have
been proposed, suggesting that the approximation should be valid if the
statistical interaction is weak.
One can, for instance, devise a criterion for validity of the MFA via a
self-consistency argument [\chewilwithal,\mccmac]. After making the MFA,
particles move in circular orbits due to the uniform magnetic field.
One can then evaluate the number of particles $Q$
contained inside a typical orbit and express it in terms of the strength of 
the statistical interaction $\alpha$ (normalized so that $\alpha=1$
transmutes bosons to fermions and vice versa). 
When $Q$ is sufficiently large, the
granularity of the statistical magnetic field is unimportant and the MFA
is deemed valid. One finds [\chewilwithal,\trug,\zee,\mccmac] 
$Q\sim \alpha^{-2}$ for fermion-based anyons and $Q\sim \alpha^{-1}$ for
boson-based anyons, yielding the following criterion of validity:
$$
\eqalign{
        \a\ll1\qquad&\hbox{boson-based anyons}\cr
        \a^2\ll1\qquad&\hbox{fermion-based anyons}\cr}
\eqn\howdy
$$
Thus, we see
that the approximation is valid near Bose statistics
for boson-based anyons, and near Fermi statistics for fermion-based anyons;
the difference in powers of $\a$ indicates that one can be slightly further
from conventional statistics for fermion-based anyons before the
approximation breaks down.

In a recent paper [\caemac], another means of justifying  the MFA which
does not rely on self-consistency was explored. The argument, alluded to in
Ref.\ [\wenzee], consists in using parity violation in the scattering of
anyons [\marwil]. This results in an asymmetry in the scattering
cross-section of two anyons, which allows one to evaluate a mean scattering
angle for a typical anyon trajectory. From this, one can calculate the mean 
radius of the resulting quasi-circular orbit and extract (as in the 
self-consistent argument described above) the number of particles $Q$
contained inside a typical orbit. The procedure, carried out for a system
of boson-based anyons, yields the same result as found in the
self-consistent argument described above, \ie,\ $Q\sim\alpha^{-1}$,
recovering the criterion for validity of the MFA, $\alpha\ll1$.

One peculiarity of the above ``asymmetric-scattering'' approach is that it
is necessary to introduce a conventional interaction between the anyons. In
[\caemac], this interaction was parameterized by introducing a phase shift
put into the lowest partial wave. The MFA result was recovered for
``generic'' phase shift, \ie, for one not near a multiple of $\pi$.

At a purely pragmatic level, this need for a conventional interaction
arises because the scattering of two free anyons does not violate parity
[\marwil], and thus the mean radius of curvature calculated as above would
be infinite.  At a more profound level, it was conjectured [\caemac] that
for free  anyons, the motion of a single anyon in an anyon gas is not
correctly analyzed by regarding it as a sequence of individual scatterings
(thus ignoring interference effects between subsequent scatterings). This
conjecture is supported by a study of scattering of a charged particle
off a semi-infinite
rectangular lattice of flux tubes [\kiewei],
wherein it was found that in the limit where the flux per anyon goes to
zero, the scattering agrees with
the motion of a particle in a uniform magnetic field
(and hence with that of the MFA).

If this conjecture is correct, it is interesting that the
asymmetric-scattering approach of [\caemac] (which did not take into
account multiple scatterings) seemed to work so well.
Apparently, the coherence found in free anyon scattering which necessitates
the inclusion of interference between subsequent scatterings is absent in
the scattering of anyons with an additional interaction.

Ideally, as suggested in [\kiewei], it would be interesting to study the
scattering of one anyon off a random array of flux tubes (more closely
approximating the motion of an anyon in a gas of anyons) to see if the
result of [\kiewei] persists.

Since such studies of multiple scatterings appear exceedingly difficult, it
is worthwhile exploring further the validity of the MFA in a situation
where the statistical interaction is supplemented by a conventional
interaction, ignoring multiple scatterings.
In this paper, we apply the asymmetric-scattering method to a
more realistic system, namely, a gas of anyons with hard-disk repulsion.
(Indeed, as we will show, parameterizing an added conventional interaction
by a phase shift in the lowest partial wave is dangerously misleading in a
rather subtle way, since
it implies that the {\sl conventional} interaction itself violates parity.)
The scattering of two such anyons exhibits clearly an 
asymmetry in the cross-section [\suzsribhalaw] due to parity
violation, allowing us to calculate a mean radius of curvature for a
typical orbit, and thus to deduce a criterion for the validity of the MFA.
In the course of the analysis,
we will present some peculiarities in the partial wave
decomposition for anyons, which sheds some light on a seemingly bizarre
situation arrived at in [\marwil].

\section{Generalities}
To begin, it is perhaps worthwhile to discuss briefly the familiar case of
conventional (non-AB) scattering, in order to set up notation, \etc\ The
relative Hamiltonian for two particles interacting via a potential
$V(\bfr)$ is
$$
H=-\half\left({\partial^2\over\partial r^2}+{1\over r}
{\partial\over\partial r}+{1\over
r^2}{\partial\over\partial\theta^2}\right)+V(\bfr).
$$
A scattering solution is sought of the form of an incident wave plus a
scattered wave:
$$
\psi(r,\theta)=e^{ikr\cos\theta}+{e^{ikr}\over\sqrt{r}}f(\theta),
$$
the scattering amplitude
$f(\theta)$ being related to the scattering cross-section in the usual
way, $d\sigma/d\theta=|f(\theta)|^2$. 

If the particles are identical bosons, the scattering amplitude is in fact
$F(\theta)=f(\theta)+f(\theta-\pi)$, whereas for fermions it is
$\tilde F(\theta)=f(\theta)-f(\theta+\pi)$.

With parity \p\ defined
as $(r,\theta)\to(r,-\theta)$, it is clear that if $V$
is an even function of $\theta$, the Hamiltonian respects \p, and both
$f(\theta)$ and the cross-section will be even functions of $\theta$. If
the potential is cylindrically symmetric,
and a partial wave
expansion can be performed:
$$
f(\theta)={1\over\sqrt{2\pi i k}}\sum_{m=-\infty}^\infty
e^{im\theta}\left(e^{2i\delta_m}-1\right).
$$
Since cylindrical symmetry implies \p\ invariance,
$\delta_{-m}=\delta_m$. This implies, among other things, that $F$ and
$\tilde F$ are sums over even and odd values, respectively,
of $m$, and also the familiar
result that fermions cannot scatter at angle $\pi/2$ (equally valid, under
normal circumstances, in two or three dimensions.) 

Indeed, these `normal circumstances' are so ubiquitous that any suggestion
that identical fermions {\sl could} scatter at angle $\pi/2$ [\marwil]
seems at first
downright blasphemous. It is perhaps worth mentioning, therefore, that with
an unusual interaction fermions can easily be made to scatter at angle
$\pi/2$: a fairly silly but perfectly valid example would be
`hard-rectangle' scattering in the classical limit, where the infinite
rectangular
potential barrier
is skewed at an angle $\pi/4$ with respect to the
relative momentum vector of the two particles. Scattering off the
long/short face would produce
a scattering angle of $\pm\pi/2$, respectively,
with the long face yielding a much larger scattering amplitude. Thus, 
the fermionic amplitude at $\pi/2$, $\tilde F(\pi/2)=f(\pi/2)-f(-\pi/2)$,
would be nonzero. Clearly,
the standard wisdom that identical fermions cannot scatter at angle $\pi/2$
applies only to parity-invariant potentials.

Although in potential scattering the ability to perform a partial wave
decomposition implies \p\ invariance, one is certainly
free to contemplate
a situation where $\delta_m$ is {\sl not} an even function
of $m$, in which case fermions {\sl can} scatter at angle $\pi/2$. As will
be seen, this is exactly what occurs when a statistical interaction is
added: one has rotational invariance, so that a partial wave
decomposition is possible, yet \p\ invariance is broken, so that in general
$\delta_m\neq\delta_{-m}$.

Pure AB scattering and the scattering of free anyons are both described by
the Hamiltonian
$$
H=-\half\left({\partial^2\over\partial r^2}+{1\over r}
{\partial\over\partial r}+{1\over
r^2}\left({\partial\over\partial\theta}-i\alpha\right)^2\right).
$$
Here, $\a$ represents the flux in AB scattering in units of the flux
quantum, and the statistical parameter in anyon scattering (normalized so
that $\a=1$ represents transmutation between fermions and bosons). Again,
one looks for a solution in the form of an incident wave plus scattered
wave. Technically, the incident wave must be a plane wave
``modulated'' by an additional
phase due to the vector potential in order to describe a uniform particle
current (specifically, $\psi_{\rm inc}=\exp i(k r\cos\theta+\a\theta)$), 
but, as has been emphasized by Hagen [\hagen], we can be cavalier
about this since a ``naive'' incident plane wave describes the correct
covariant particle current as $r\to\infty$, and, indeed, the scattering
amplitude is unaffected by the choice of incident wave except in the form
of a $\delta$-function in the
forward direction. In fact, in what follows we will simplify life by
ignoring contributions to $f(\theta)$ which represent a $\delta$-function in
the forward direction.

The solution of this problem is well known [\ahaboh,\hagen]: 
the phase shifts, scattering
amplitude and differential cross-section for AB scattering are
$$
\eqalign{
	\delta^{AB}_m(\a)&={\pi\over2}\left(|m|-|m-\a|\right),\cr
	f^{AB}_\a(\theta)&=-{\sin\a\pi\over\sqrt{2\pi i k}}
		{e^{i(N+1/2)\theta}\over\sin\theta/2},\cr
	{d\si\over d\theta}^{AB}&=
		{\sin^2\a\pi\over2\pi k\sin^2\theta/2}.\cr}
\eqn\aharbohm
$$
where $N$ is the integer part of $\a$, while for boson-based anyon
scattering,
$$
\eqalign{
	F^{AB}_\a(\theta)=f^{AB}_\a(\theta)+f^{AB}_\a(\theta+\pi)
		&=-{\sin\a\pi\over\sqrt{2\pi i k}}
		{2e^{i(N+1/2)\theta}e^{i(-)^N\theta/2}\over\sin\theta},\cr
	{d\Sigma_\a\over d\theta}^{AB}=|F^{AB}_\a(\theta)|^2&=
		{2\sin^2\a\pi\over\pi k\sin^2\theta},\cr}
$$
It is interesting to observe that, in spite of the fact that the
Hamiltonian is not \p\ invariant, the scattering cross-section is.

One important property of $f^{AB}_\a(\theta)$ is its
periodicity (up to a
phase) under an integer change of $\a$. 
This is due to the fact that only the fractional part of the 
flux is physically relevant in AB scattering. Thus, the differential
cross-section must be invariant under $\a\to\a+n$, although the scattering
amplitude itself can (and does) change by a phase: one finds
$$
f^{AB}_{\a+n}(\theta)=(-)^ne^{i n\theta/2}f^{AB}_\a(\theta).
\eqn\dog
$$
This relation, expressed in terms of phase
shifts, is
$$
\delta^{AB}_m(\a+n)=\delta^{AB}_{m-n}(\a)-{n\pi\over2}.
\eqn\cat
$$

Now let us suppose a cylindrically symmetric potential $V(r)$ 
is added, and study scattering as $\a$ is varied, holding the potential
fixed. Again
we can perform a partial wave expansion, obtaining phase shifts
$\Delta_m^V(\a)$ which, in principle, can be determined in terms of the
potential. These phase shifts obey the same periodicity requirement as the
$\delta^{AB}_m(\a)$:
$$
\Delta^{V}_m(\a+n)=\Delta^{V}_{m-n}(\a)-{n\pi\over2}.
\eqn\fish
$$

It is useful to separate $\Delta_m^V$ into the AB phase shift and a
residual one due to the potential:
$$
\Delta^{V}_m(\a)=\delta^{AB}_{m}(\a)+\delta^{V}_{m}(\a).
$$
The key observation is that $\delta^V_m$, although due to the potential
{\sl must necessarily depend upon} $\a$. This can be seen most easily
by substracting the
periodicity conditions \cat\ and \fish; one obtains
$$
\delta^{V}_{m}(\a+n)=\delta^{V}_{m-n}(\a).
\eqn\period
$$

Although convenient, this separation is not without its dangers. For
instance, in 
Refs.\ \caemac,\marwil, a ``conventional'' interaction between anyons was
parameterized by an additional phase shift $\delta$ in the lowest partial
wave $m=0$, and \p\ violation in the resulting scattering as $\a$ varied
was discussed. However,
keeping the residual phase shifts $\delta_m^V$ fixed as $\a$ varies
implies that the conventional
interaction must evolve as $\a$ changes, in a rather complicated way.
Indeed, if the added phase
shifts are an even function of $m$, the added interaction is \p\ invariant
at $\a=0$, but away from this point the conventional
interaction (suitably
evolved so as to keep the added phase shifts constant) necessarily
violates parity. 

This can be clearly demonstrated in a simple way for
integer values of $\a$, as follows. As stated above, for conventional
scattering ($\a=0$), \p\ invariance implies that $\delta_m$ is an even
function of $m$; in the present context, thus,
$\delta_{-m}^V(0)=\delta_{m}^V(0)$. The periodicity requirement \period\ 
of $\delta_m^V(\a)$ then implies the following ``shifted evenness''
requirement
for any \p-conserving potential at nonzero integer values of $\a$:
$$
\delta^V_{n+m}(n)=\delta^V_{n-m}(n)
\eqn\shifted
$$
The hypothetical potential which produces $\a$-independent phase shifts
will certainly not obey \shifted, since its phase shifts are an even
function of $m$. (It takes but a moment to convince oneself that phase
shifts which are even both about $m=0$ and about $m=n\neq0$ are pathological
beyond all reason.) Thus, the potential must evolve to a
parity-violating one for nonzero integer $\a$. For noninteger $\a$, while
we have not proven that the potential must evolve to one which
violates parity if the residual phase
shifts are independend of $\a$, continuity arguments make it exceedingly
plausible.

To make these ideas concrete, we can examine the case discussed in
[\caemac,\marwil], where the potential was parameterized by a phase shift
in the lowest partial wave. 
At $\a=0$, such a phase shift is the result of
a rather peculiar $\delta$-function potential
if the other phase shifts are truly zero. A complete discussion would
involve delving into the subject of self-adjoint extensions
[\selfadj], but for our purposes a less mathematical approach
will suffice [\pedest]. Consider scattering in the presence of
a cylindrical potential well of radius $a$ and depth $\la/\pi a^2$, in
the limit $a\to0$, whence the potential tends towards $\la\delta^2(\bfr)$.
This ``regularized'' $\delta$-function enables us to adopt the physically
reasonable boundary condition that the wave function is finite at the
origin (mathematically, this is equivalent to making a particular
choice of self-adjoint extension) [\regular]. Keeping the
momentum $k$ fixed, it is easy to show that, as $a\to0$, the phase shift in
the lowest partial wave goes to zero as $1/\log ka$. However, if we
consider the coefficient $\la$ to be itself a function of $a$, which goes
to zero as $a\to0$, then the lowest partial wave is given by
$$
\tan\delta_0\simeq-{\pi\over2(\log{ka\over2}+\ga)+{4\pi\over\la}}.
\eqn\tangent
$$
This can be made nonzero if $\la$ goes to zero just quickly enough to
cancel the divergent logarithm; specifically, if
$$
\la\to{4\pi\over D-2\ga-\log{k^2r^2\over4}}
$$
where $D$ is a constant, then
$$
\tan\delta_0\to-{\pi\over D}.
\eqn\xxx
$$
All other partial waves will be zero, essentially because the centrifugal
barrier prevents them from coming into contact with the potential.

It is a straightforward matter now to check how the phase shifts evolve
when we turn on the statistical interaction; not surprisingly, one finds
that $\delta_0$ drops to zero immediately once $\alpha$ becomes nonzero:
the delicate balance between the two divergent terms in the denominator of
\tangent\ is lost.
Furthermore, all the phase shifts remain zero until $\alpha=1$, at which
point the first partial wave is nonzero and all others are zero, in
agreement with \period.

Since the initial potential does not maintain constant phase shifts for
nonzero $\a$, some other potential is required. It would be interesting to
find this potential, but we have not managed to do so. (It would certainly
be very unusual, since, according to the general arguments given above, we
expect it to violate parity.)

A second concrete example of the above ideas is the example of hard-disk
scattering [\suzsribhalaw], which is discussed below, and with which we
will examine the question of justification of the MFA for the anyon gas.

\section{Anyons with hard-disk repulsion}

Consider the case of the scattering
of two anyons whose interaction is described by a hard-disk repulsion 
within a region of size $a$ [\suzsribhalaw]. Parity violation in the
scattering cross-section will enable us to use the ideas presented in Ref.\ 
[\caemac] to address the question of the validity of the MFA.

The scattering amplitude $f_\a(\theta)$ can be determined in a rather
conventional fashion; as mentioned above, it is useful to separate it into
a pure AB piece, $f_\a^{AB}(\theta)$, and a new piece which owes its
existence due to the hard-disk interaction:
$$
f_\a(\theta)=f_\a^{AB}(\theta)+f_\a^{HD}(\theta)
$$
The latter term can be written in terms of
phase shifts $\delta_m^{HD}(\alpha)$, which are computed by
imposing the boundary condition $\psi(a)=0$; explicitly, one finds
$$
\tan\delta_m^{HD} = \tan\delta_{|m-\alpha|}
={J_{|m-\alpha|}(ka)\over N_{|m-\alpha|}(ka)}.
$$
These phase shifts clearly
depend on the statistical parameter
$\alpha$, and, moreover, satisfy the periodicity requirement \period, in
accord with the discussion of the previous section.
In fact, at low energies the tangent of the phase shifts have the following
limiting behaviour:
$$
\tan\delta_m^{HD} \to\beneathrel{ka\to0}\under\longrightarrow
\cases{	(ka)^{2|m-\a|}&$m-\a\neq0$\cr
	\log ka&$m-\a=0$\cr}.
\eqn\limiting
$$
If $\a=0$, the particles are normal bosons and the
dominant partial wave is $m=0$; the potential thus gives approximately the
scattering behaviour discussed in [\marwil,\caemac].
As $\a$ evolves towards 1, the $m=1$
partial wave starts to contribute significantly and is, indeed, the
dominant one beyond $\a=1/2$. When $\a=1$ we have a shift by one unit of
the phase shifts compared with $\a=0$, in agreement with the previous
section.

The hard-disk part of the scattering amplitude is
$$
f_\a^{HD}(\theta)={1\over \sqrt{2\pi ik}}\sum_{m=-\infty}^\infty
e^{im\theta}e^{i\pi(|m|-|m-\alpha|)}(e^{2i\delta_m^{HD}(\alpha)}-1).
\eqn\harddisk
$$

This problem is a natural situation in which we can apply the 
method presented in Ref.\ \caemac\ in order to check the validity of the MFA.
We start by considering boson-based anyons. The cross-section for the
scattering of two anyons is given by 
$$
{d\Sigma_\a\over d\theta}
=\vert f_\a(\theta)+f_\a(\theta -\pi)\vert^2.
$$
for which an even integer $\alpha$ describes bosons while an odd one
describes fermions.

Our goal is to compute a mean scattering angle, and from it to compute a
mean radius of curvature for a typical anyon trajectory.
In order to proceed with the argument, we now have to evaluate the quantity
$$
X\equiv\left|\int d\theta\,\theta{d\Sigma_\a\over d\theta}\right|,
\eqn\meanangle
$$
which, once normalized by the total cross-section $\Sigma$, gives the mean
radius 
of curvature $\bar\theta =X/\Sigma$. The mean length between scatterings is
given by $L=1/s\Sigma$ where $s$ is the anyon density. These 
two ingredients enable us to evaluate, if the mean scattering angle is small, 
the mean radius of curvature of the quasi-circular orbit, which is given by
$$
\bar R={L\over\bar\theta}={1\over sX}.
$$ 
The number of particles contained whithin this roughly-circular orbit will 
then be
$$
Q\sim {\bar R}^2={k^2\over sX'^2}
$$
where we have defined a new quantity $X'=kX$, which depends only 
on the parameters $\alpha$ and $ka$. 

The momentum $k$ can be estimated in the following
way. The mean radius $\bar R$, which classically represents a length scale for 
the particle's trajectory, can be used to define at the quantum level the size 
of the region where the wave function is localized. The uncertainity principle
then gives us a minimum momentum of $k\sim 1/\bar R=sX'/k$. The number of
particles $Q$ can then be expressed in the simple form
$$
Q\sim{1\over X'}.
\eqn\cairo
$$
The MFA is declared valid if $Q\gg 1$, since that
implies that the graniness of the distribution of particles is on a much
smaller scale than the length scale of the particles' trajectories. This
gives us the criterion of validity
$$
X'< X'_c,
\eqn\criter
$$ 
where $X'_c\ll1$, the appropriate value being dictated by physical
considerations.

From the definition of the mean angle 
\meanangle\ and using the scattering amplitudes \aharbohm\ and \harddisk,
we can compute the cross-section $d\Sigma_\a/d\theta$, and thus
the quantity $X'$. We find the following rather unwieldy expression:
$$
\eqalign{
	X'=&8\sum_{l\not=l'}(-)^{l+l'\over 2}
	{1\over l-l'} \Bigl[{J_{+}J'_{+}(J_{+}N'_{+}-J'_{+}N_{+})
	\over(N_{+}^2+J_{+}^2)(N^{'2}_{+}+J^{'2}_{+})}
	-{J_{-}J'_{-}(J_{-}N'_{-}-J'_{-}N_{-})\over 
	(N_{-}^2+J_{-}^2)(N^{'2}_{-}+J^{'2}_{-})}\Bigr] \cr
	+&8\sum_{l,l'\neq(0,0)}
	(-)^{l+l'\over 2}{2\over l+l'}{J_{-}J'_{+}\over
	(N_{-}^2+J_{-}^2)(N^{'2}_{+}+J^{'2}_{+})}\cr
	&\qquad\qquad\quad\times\bigl[(N_{-}N'_{+}+J_{-}J'_{+})2sc+
	(J_{-}N'_{+}-J'_{+}N_{-})(c^2-s^2)\bigr] \cr
	+&{8\over\pi}\sum_{l}
	\Bigl\{J_{-}{N_{-}s^2+J_{-}sc\over N_{-}^2+J_{-}^2}I_{l-2}
	-J_{+}{N_{+}s^2+J_{+}sc\over N_{+}^2+J_{+}^2}I_l\Bigl\},\cr }
\eqn\monster
$$
where the sums are over $l,l' = 0, 2, 4,\dots$, and
where we have used the notation $J_{\pm}=J_{l\pm\alpha}(ka)$,
$J'_{\pm}=J_{l'\pm\alpha}(ka)$, $s=\sin(\alpha\pi)$ and
$c=\cos(\alpha\pi)$. Furthermore,
$I_l$ is given by
$$
I_l=\pi\sum_{k=1}^{l/2}{(-1)^k\over k} +\pi\ln 2.
$$
\monster\ is sufficiently ugly that we must turn to a numerical analysis
in order to extract the behaviour of $X'$ as we vary
the parameters $\alpha$ and $ka$.
The resulting graphs are then visually  analysed to determine the 
regions in the parameters $\alpha$ and $ka$ which satify the
criterion of validity of the MFA \criter. Without loss of generality, we
can restrict the range of $\a$ to
$[0,1]$, $\alpha=0$ and $\alpha=1$ describing respectively bosons and
fermions.

Fig.\ 1 presents a three-dimensional plot of $X'$ in terms of the parameters 
$\alpha$ and $ka$.
Fig.\ 2 diplays a contour plot of the same function where the
solid region indicates the values of $\alpha$ and $ka$ satisfying the criterion 
of validity \criter\ (where we have chosen $X'_c=0.1$).
Fig.\ 3 shows a plot of $X'$ in terms of $\alpha$ for fixed $ka$ (which has been
put at the value $ka=0.1$).    

By inspection of the plots diplayed in Fig.\ 1 
and Fig.\ 2, we are able to conclude
that the criterion of validy is assured in three different limits in the
parameters: 
1) $\alpha$ near 0, i.e. anyons near the bosonic regime; 
2) $\alpha$ near 1, i.e. anyons near the fermionic regime;
3) $ka$ very small for $\alpha$ arbitrary.
The last condition is not very surprising 
since as $ka$ goes to zero the particles become free anyons, whose
scattering does not
violate parity [\marwil]. The trajectory of an
anyon is then apparently straight
(on average) and the argument based on parity violation described above
loses is
utility for studying the validity of the MFA.

The plot displayed in Fig.\ 3 
allow us to go much further in our analysis of the
MFA since it shows not only the regions in the parameter
$\alpha$ where the criterion of validity \criter\ is respected,
but also gives an
indication of the dependence of $X'$ on $\a$ near $\a=0,
1$. As we can see,
one can expect $X'$ to depend linearly on the 
statistical parameter $\alpha$ near the bosonic regime and quadraticaly on the
statistical parameter $\tilde\alpha=1-\alpha$ (measuring the departure from
Fermi statistics) near the fermionic regime.

This can be understood analytically as follows. We are interested primarily
in values of $ka$ considerably less than 1, since on the one hand the
momentum satisfies $k\bar R\sim1$, and on the other hand the mean radius of
curvature $R$ should be greater than $a$ in order for the whole approach of
considering isolated scatterings of anyons to be reasonable.
In the bosonic regime, $\alpha$ near 0, $X'$ has the limiting form
$$
X' \to {2\pi^3\log 2\over (\gamma +\log(ka/2))^2 +\pi^2}\alpha,
\eqn\bosonlimit
$$
which clearly shows a linear dependence on the statistical parameter $\alpha$. 
The MFA is, therefore, deemed justified for
$$
\alpha \ll 1.\qquad\hbox{near bosons}
\eqn\hattrick
$$
In the fermionic regime, $\tilde\alpha$ near 0, 
the situation is more complicated. Writing $ka=\eps$, for $\tilde\a\ll1$,
we find

$$
X' \to \pi^3\eps^2\tilde\a(c\eps^2+d\tilde\a),
\qquad c,d\sim1.
\eqn\fermionlimit
$$
Thus, for $\tilde\a\ll\eps^2$, the dependence of $X'$ on $\tilde\a$ is
linear, while for $\tilde\a\gtwid\eps^2$, it is quadratic. However, we are
concerned with the dependence when $X'$ is near its critical value $X'_c$.
For $\eps$ considerably smaller than 1, the appropriate dependence on
$\tilde\a$ can easily be seen from \fermionlimit\ to be quadratic,
indicating that the MFA is justified if
$$
\tilde\alpha^2 \ll 1.\qquad\hbox{near fermions}
\eqn\nearfer
$$
In both cases, we recover the results obtained previously
[\chewilwithal,\trug,\zee,\mccmac], but perhaps in a more trustworthy way
since self-consistency was not a part of our argument.

In summary, we have discussed general features of Aharonov-Bohm scattering
with an additional conventional interaction, with a particular emphasis on
the parity violation which may be found in the scattering cross-section.
The danger of parameterizing the added interaction by phase shifts was
pointed out; the essential point is that the added phase shifts must depend
on the statistical parameter. In the latter part of the paper, we
applied the ``asymmetric-scattering'' method discussed in [\caemac] on
anyons with a hard-disk repulsion in order to extract a criterion for the
validity of the MFA. In agreement with previous works, we found that the
statistical interaction must be weak both for anyons near bosons and near
fermions, the quantitative criteria being given by \hattrick\ and \nearfer.

We thank T. Gisiger for help with the numerical work, and F. Wilczek for
useful discussions.
This work was supported in part by the Natural Science and
Engineering Research Council of Canada and the Fonds F.C.A.R. du Qu\'ebec.

\refout

\vskip\chapterskip
\line{\fourteenrm\hfil FIGURE CAPTIONS\hfil}\vskip\headskip
\sfcode`\.=1000 \interlinepenalty=1000

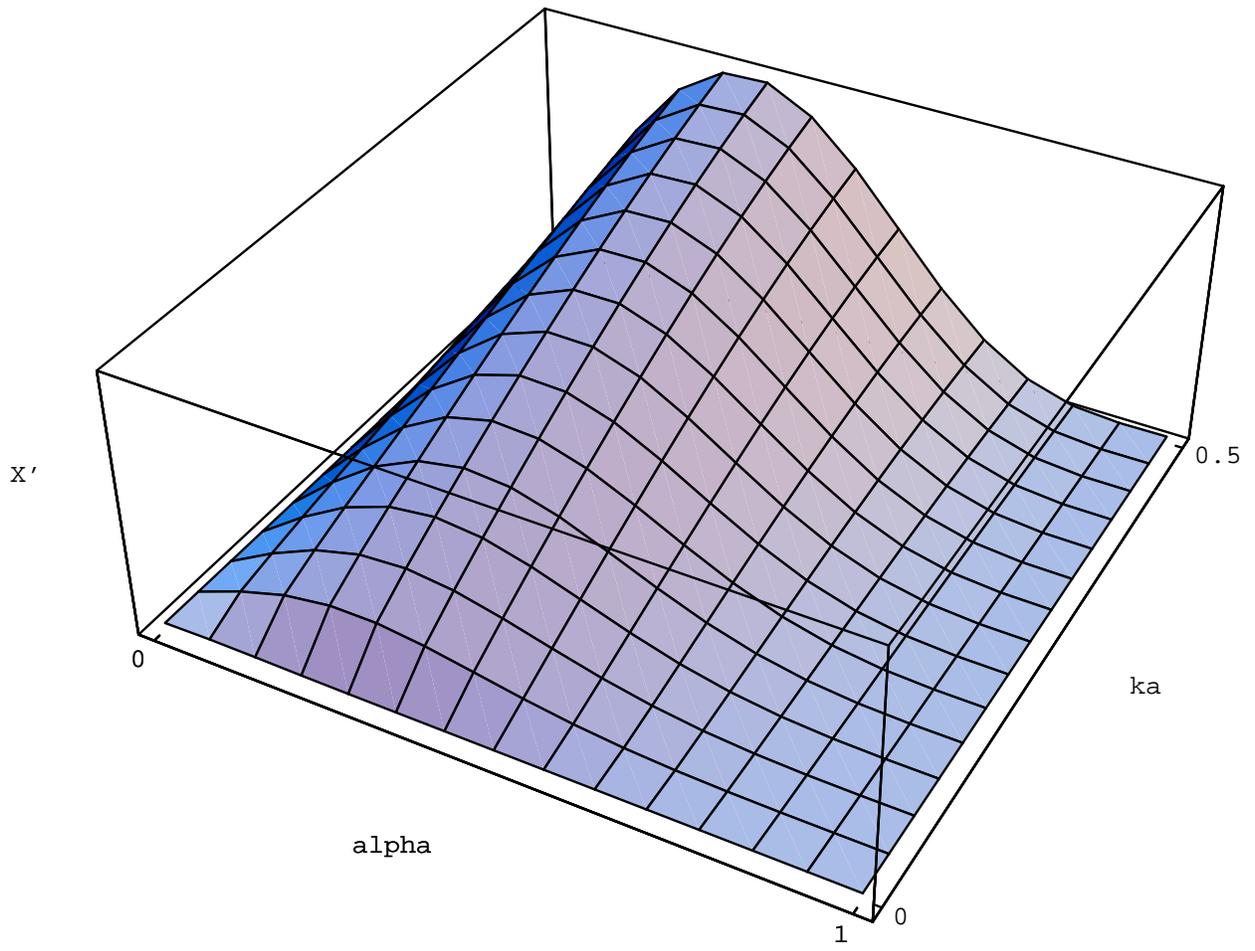
\noindent{FIG. 1.} Three-dimensional plot of $X'$ in terms of the parameters
$\alpha$ and $ka$.

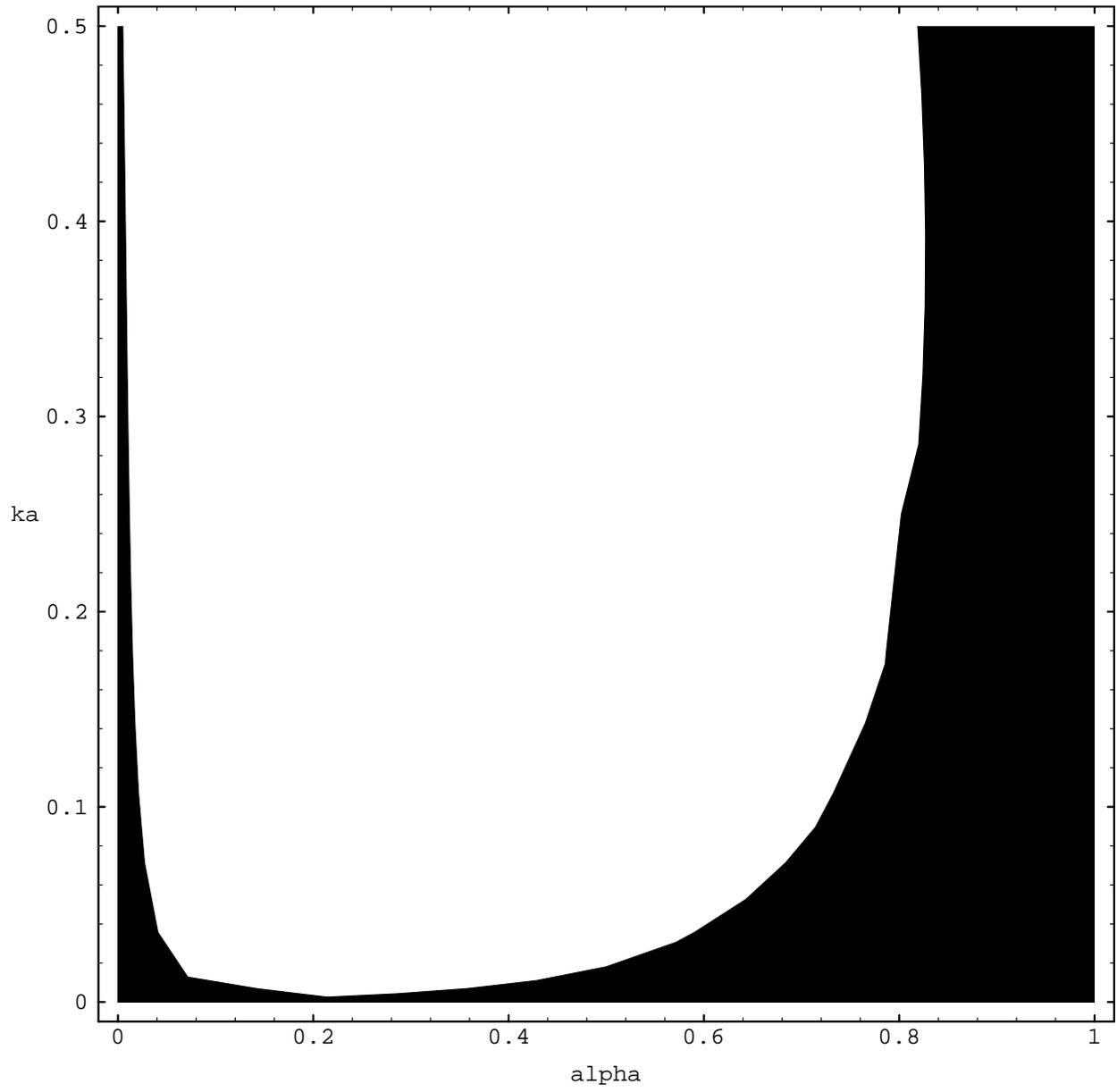
\noindent{FIG. 2.} Contour plot of $X'$ in terms of the parameters
$\alpha$ and $ka$. The solid region indicates the values of $\alpha$ and 
$ka$ satisfying the criterion of validity \criter, with $X'_c=0.1$.

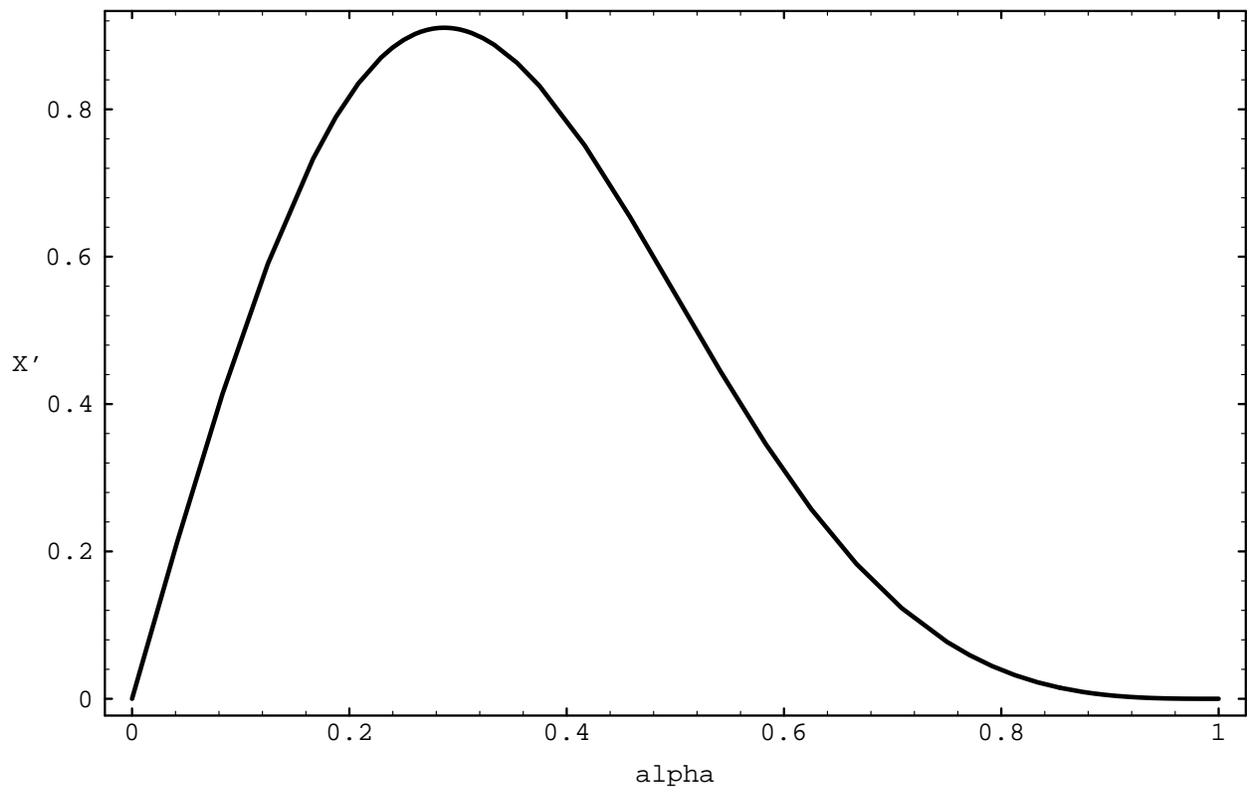
\noindent{FIG. 3.} Plot of $X'$ in term of $\alpha$ for fixed $ka$ 
(which has been put at the value $ka=0.1$).

\end